\documentclass[11pt,preprint]{aastex}

\newcommand{\thisgrb}{GRB~010222}
\newcommand{\rkc}{$R_{\rm C}$}

\slugcomment{accepted by The Astrophysical Journal Letters}

\shortauthors{Jha et al.}
\shorttitle{Redshift of \thisgrb}

\begin{document}
 
\title{The Redshift of the Optical Transient Associated with \thisgrb}

\author{Saurabh Jha\altaffilmark{1}, 
Michael A. Pahre\altaffilmark{1}, 
Peter M. Garnavich\altaffilmark{2}, 
Michael L. Calkins\altaffilmark{1}, 
Roy E. Kilgard\altaffilmark{1}, 
Thomas Matheson\altaffilmark{1}, 
Jonathan C. McDowell\altaffilmark{1}, 
John B. Roll\altaffilmark{1},
and Krzysztof Z. Stanek\altaffilmark{1}} 
\email{sjha, mpahre, pgarnavich, mcalkins, rkilgard, tmatheson,
jmcdowell, jroll, kstanek@cfa.harvard.edu}

\altaffiltext{1}{Harvard-Smithsonian Center for Astrophysics, 60
Garden Street, Cambridge MA 02138}
\altaffiltext{2}{Dept. of Physics, University of Notre Dame, 225
Nieuwland Science Hall, Notre Dame IN 46556}

\begin{abstract}
The gamma-ray burst (GRB) 010222 is the brightest GRB detected to date
by the BeppoSAX satellite.  Prompt identification of the associated
optical transient (OT) allowed for spectroscopy with the Tillinghast
1.5~m telescope at F.~L.~Whipple Observatory while the source was
still relatively bright ($R \simeq 18.6$~mag), within five hours of
the burst.  The OT shows a blue continuum with many superimposed
absorption features corresponding to metal lines at $z = 1.477$,
$1.157$, and possibly also at $0.928$.  The redshift of \thisgrb\ is
therefore unambiguously placed at $z \geq 1.477$.  The high number of
\ion{Mg}{2} absorbers and especially the large equivalent widths of
the \ion{Mg}{2}, \ion{Mg}{1}, and \ion{Fe}{2} absorption lines in the
$z=1.477$ system further argue either for a very small impact
parameter or that the $z = 1.477$ system is the GRB host galaxy
itself.  The spectral index of the OT is relatively steep, $F_\nu
\propto \nu^{-0.89 \pm 0.03}$, and this cannot be caused by dust with
a standard Galactic extinction law in the $z = 1.477$ absorption
system.  This spectroscopic identification of the redshift of
\thisgrb\ shows that prompt and well-coordinated followup of bright
GRBs can be successful even with telescopes of modest aperture.
\end{abstract}


%

\section{Introduction \label{introduction} }

Since the discovery of gamma-ray bursts
\citep*[GRBs;][]{klebesadel73}, their nature has proved enigmatic.
The CGRO and BATSE observations demonstrated they were isotropically
distributed on the sky \citep{meegan92}, which could either have been
explained by Galactic \citep{lamb95} or cosmological
\citep{paczynski95} spatial distributions.  The BeppoSAX satellite
\citep{bepposax} contributed the breakthrough in this field by
providing rapid localizations of the X-ray afterglow of a GRB to a
precision of several arcminutes.  Such precision allowed for rapid
optical \citep{paradijs97} and radio \citep{frail97} identifications
of transients associated with individual GRBs.  The identification of
the optical transient (OT) associated with GRB~970508
\citep{bond97, djorgovski97} led to the first optical spectroscopic redshift
determination for a GRB, placing it at $z \geq 0.835$
\citep{metzger97}, and thus firmly at a cosmological distance.

Despite such rapid progress and intensive followup campaigns at many
wavelengths, only some GRBs have had associated X-ray, optical or radio
afterglows, and all of these events have been among the
``long-duration'' GRB population \citep{kulkarni00}. Of these, only
$\sim$15 have had unambiguous spectroscopic redshift identification
\citep*[see those to date tabulated by][]{bloom01a}. The primary
difficulty is the combined delays imposed by the time necessary to
improve the X-ray localizations, interruption of a telescope observing
program to obtain optical imaging data, reduction and analysis of those
data relative to POSS images to identify the OT, and interruption of
another telescope observing program to obtain a spectrum (usually on a
different telescope, since there are very few combined imaging and
spectrograph instruments). By the time a spectrum is taken, the OT has
usually faded significantly, generally limiting such followup
observations to only the largest available ground-based optical
telescopes or the Hubble Space Telescope. Furthermore, observational
conditions may not be optimal immediately after a GRB event, such that
only approximately half of the known GRB redshifts were obtained with
spectroscopy of the OT itself---the others were obtained at later times
from the presumptive host galaxy, after the OT had faded.

Here we report the rapid identification of an OT associated with
\thisgrb, and the spectroscopy of the source within five hours after
the GRB occured. The quick localization of the X-ray and optical
transients allowed for spectroscopic observations on a modest 1.5~m
telescope, the smallest aperture telescope to measure the redshift of
a GRB to date.

\section{Data \label{observations} }

\begin{figure}
\epsscale{1.0}
\plotone{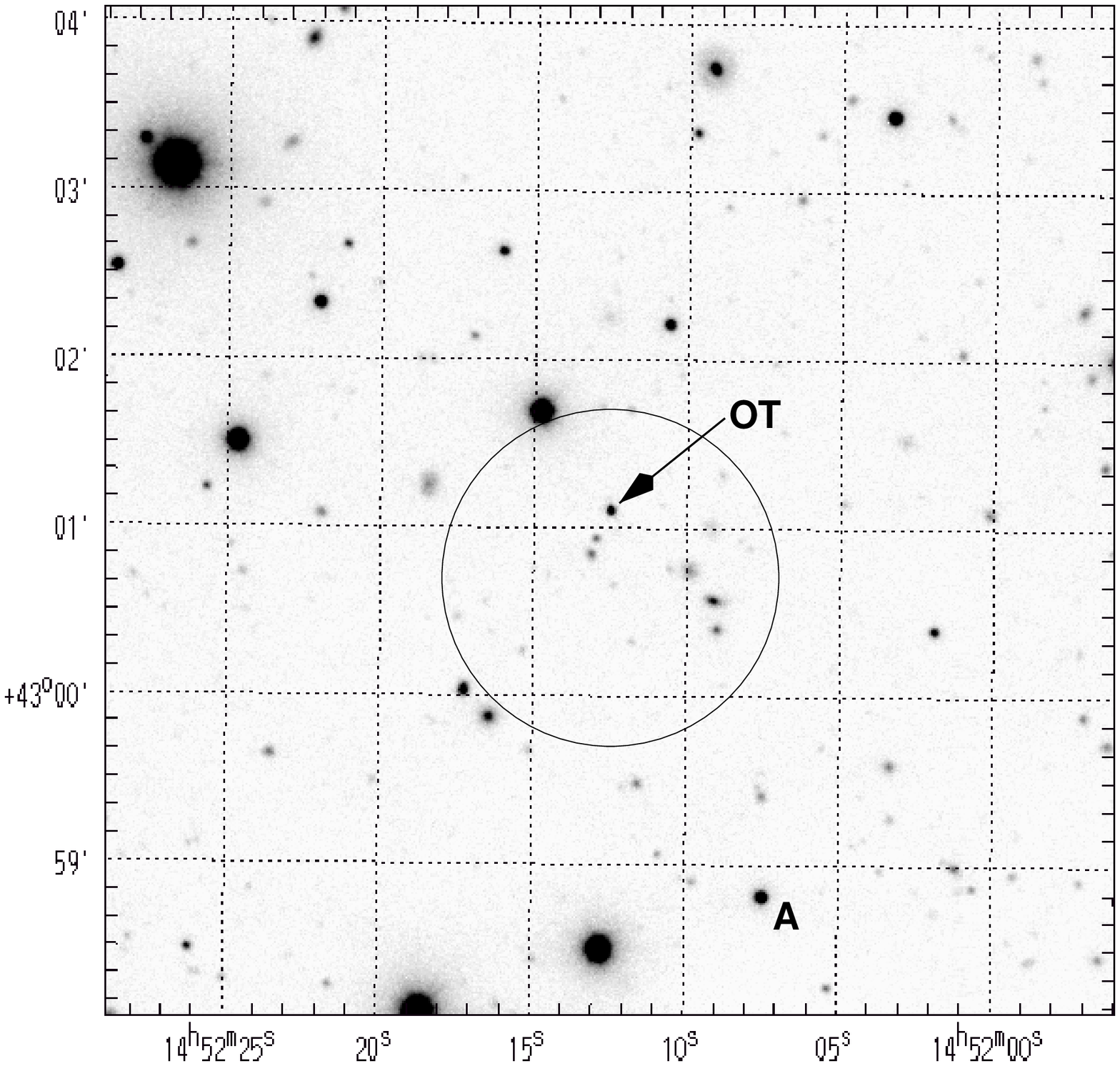}
\caption[Discovery Image of \thisgrb\ ]{Identification of the optical
transient associated with GRB~010222.  This 600s \rkc-band image,
taken with the F.~L.~Whipple Observatory 1.2-m telescope (+ 4Shooter)
beginning at UT 2001 February 22 12:12, shows the optical transient
(OT) and comparison star A of \citet{mcdowell963}. The image covers 6
arcmin on a side, with north up and east to the left. The position
error circle from the BeppoSAX followup of the X-ray afterglow
\citep{gandolfi966} is also shown. \label{fig-finding}}
\end{figure}

\thisgrb\ was detected by the Gamma-Ray Burst Monitor and Wide Field
Camera 1 instruments on board BeppoSAX at 07:23:30 UT on 2001 February
22 \citep{piro959}. An optical transient (OT) associated with \thisgrb\
was reported within several hours by
\citet{henden961}. \citet{mcdowell963} provided independent confirmation
of the OT from images taken with the F.~L.~Whipple Observatory (FLWO)
1.2~m telescope and 4Shooter CCD mosaic camera \citep{szent01}.  The
identification of the OT ($\alpha$ = $14^{\rm h}52^{\rm m}12\fs54$,
$\delta$ = $+43\arcdeg01\arcmin06\farcs2$, J2000.0) on these latter
discovery images is shown in Figure~\ref{fig-finding}.


\begin{figure}
\epsscale{1.0}
\plotone{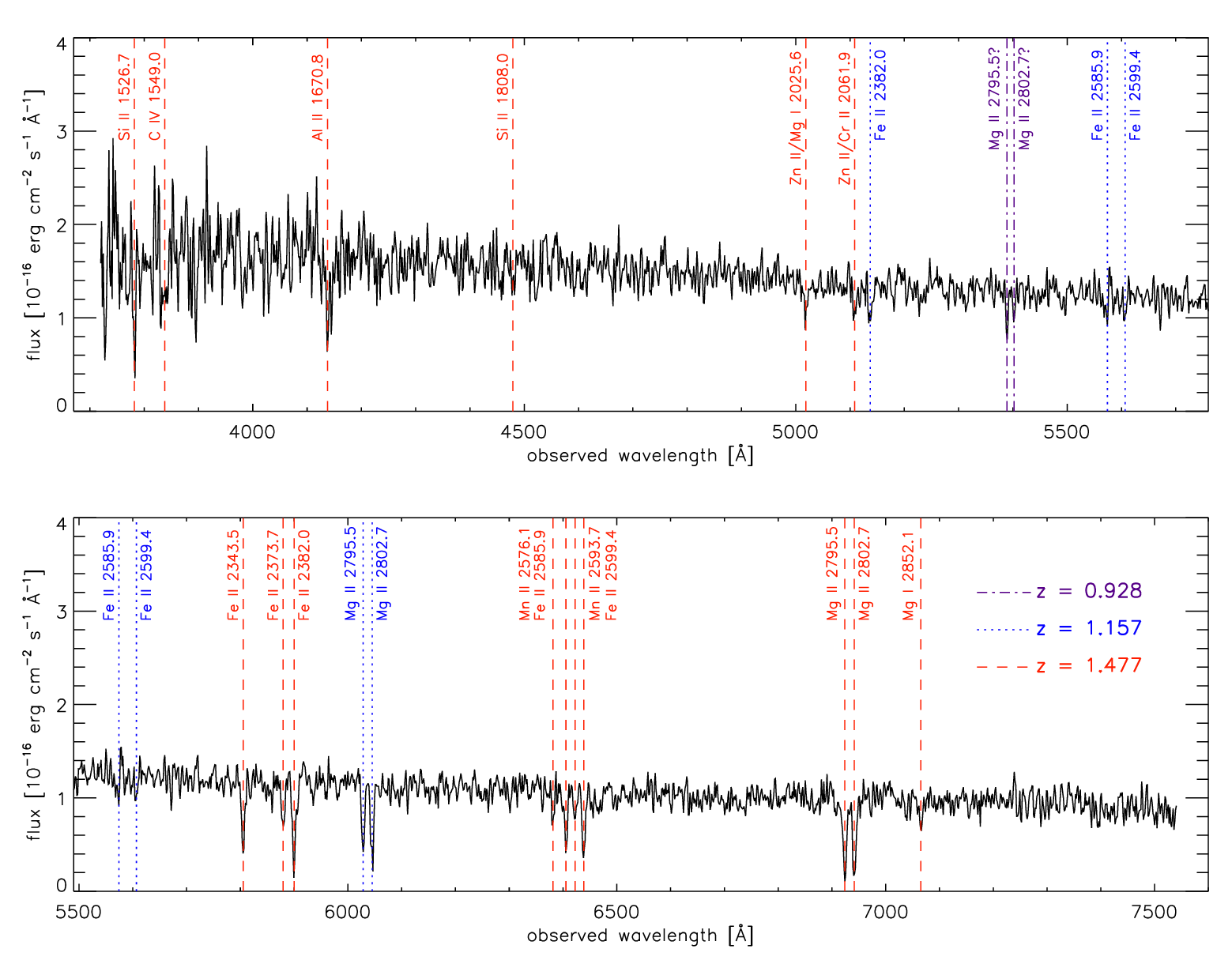}
\caption[Discovery Spectrum of \thisgrb\ ]{Discovery spectrum of the
optical transient associated with \thisgrb\ taken with the
F.~L.~Whipple Observatory 1.5-m telescope (+ FAST spectrograph) on UT
2001 February 22, approximately 5~hours after the GRB.  The OT has a
blue continuum spectrum with two absorption systems at $z=1.157$
(dotted or blue vertical lines) and $1.477$ (dashed or red vertical
lines).  Various redshifted metal absorption features are labelled.
Two absorption lines at $\lambda = 5389.1$ and
$5402.2$~\AA~(dash/dotted or purple vertical lines) may be due to an
additional \ion{Mg}{2} system at $z = 0.928$. The redshift of
\thisgrb\ is therefore constrained to lie at $z \geq 1.477$.  If the
highest-redshift absorption lines correspond to the host galaxy of the
OT, then the redshift of \thisgrb\ is $z =
1.477$. \label{fig-spectrum1}}
\end{figure}

A spectrum of the OT was obtained with the FLWO 1.5~m Tillinghast
telescope beginning at UT 2001 February 22 12:18, 4.92 hours after the
burst.  The OT apparent magnitude was $R \simeq 18.6$~mag around the
time of the spectroscopy \citep{mcdowell963,henden987}.  Despite the
faintness of the OT, it was visible on the telescope acquisition
camera, such that it could unambiguously identified and placed on the
spectrograph slit.  The observations were made with the FAST
spectrograph \citep{fast} using a 3~arcsec wide slit and 300 l/mm
grating, yielding 6 \AA\ FWHM resolution over the range $3720 <
\lambda < 7540$ \AA.  Two 1200s exposures were taken with the slit
rotated to the parallactic angle (and moreover the airmass was
$\leq$1.04), reduced in the standard manner with an optimal extraction
\citep{horne86}, and combined. Wavelength calibration was provided via
HeNeAr lamp spectra taken immediately after the OT exposures, with
minor adjustment based on night sky lines in the OT frames. We
corrected for telluric lines \citep{wade88} and flux-calibrated the
spectra with exposures of the spectrophotometric standard star Hiltner
600 \citep{stone77}, also taken at the parallactic angle, yielding
relative fluxes in our OT spectrum accurate to $\sim$5\% over the
observed wavelength range.  The discovery spectrum is shown in
Figure~\ref{fig-spectrum1}, and preliminary results from it have been
reported previously \citep{garnavich965,jha974}.  Four additional
spectra of $1800$~s each were obtained the following night (UT 2001
February 23) when the OT had faded by $\sim$1.5 to 2~mag
\citep{stanek01}.  The dispersed spectrum was at the detection
threshold of the spectrograph, hence no reliable results could be
obtained from these data.


\section{Results \label{results}  }

The optical spectrum of \thisgrb\ in Figure~\ref{fig-spectrum1} shows
a blue continuum that is typical for GRBs.  Superimposed upon this
continuum are a number of strong and weak absorption line systems at
$z = 1.157$ and $1.477$, which are identified by the metallic lines of
\ion{Mg}{1}, \ion{Mg}{2}, \ion{Fe}{2}, \ion{Mn}{2}, \ion{Si}{2},
\ion{Al}{2}, \ion{Zn}{2}, \ion{Cr}{2}, and \ion{C}{4}.  Two additional
lines are tentatively identified with \ion{Mg}{2} at $z = 0.928$
because these lines are weaker and no other lines are found at a
similar redshift (although the S/N is worse at these shorter
wavelengths).  All three systems were independently detected by
\citet{bloom01b} and confirmed by \citet{castro01} from spectroscopy
at the Keck Observatory.  Our line identifications are summarized in
Table~\ref{table-line-IDs}; as is typical for metal-line absorption
systems, the \ion{Mg}{2} lines are the strongest in the spectrum,
nearly reaching zero flux for the highest redshift system even at this
relatively low spectral resolution.

\begin{deluxetable}{ccccc}
\tablecaption{Absorption Line Identifications in the Spectrum of
GRB~010222. \label{table-line-IDs}}
\tablewidth{0pt}
\tablecolumns{5}
\tablehead{
\colhead{Observed} & \colhead{Line} & \colhead{Rest} &
\colhead{Rest-frame} & \colhead{Redshift} 
\\
\colhead{Wavelength} & \colhead{Identification}	& \colhead{Wavelength}
& \colhead{Equivalent Width} & \colhead{ } \\
\colhead{[\AA]}	& \colhead{ } & \colhead{[\AA]} & \colhead{[\AA]} &
\colhead{ }
} 
\startdata
7065.4  &      \ion{Mg}{1}              & 2852.1 & 0.9 $\pm$ 0.2 & 1.477 \\ 
6941.7  &      \ion{Mg}{2}              & 2802.7 & 2.7 $\pm$ 0.2 & 1.477 \\ 
6924.0  &      \ion{Mg}{2}              & 2795.5 & 3.0 $\pm$ 0.2 & 1.477 \\ 
6438.5  &      \ion{Fe}{2}              & 2599.4 & 1.9 $\pm$ 0.2 & 1.477 \\ 
6422.4  &      \ion{Mn}{2}              & 2593.7 & 0.7 $\pm$ 0.2 & 1.476 \\ 
6405.4  &      \ion{Fe}{2}              & 2585.9 & 1.5 $\pm$ 0.2 & 1.477 \\ 
6381.2  &      \ion{Mn}{2}              & 2576.1 & 0.7 $\pm$ 0.2 & 1.477 \\ 
5900.1  &      \ion{Fe}{2}              & 2382.0 & 2.4 $\pm$ 0.2 & 1.477 \\ 
5879.6  &      \ion{Fe}{2}              & 2373.7 & 1.2 $\pm$ 0.2 & 1.477 \\ 
5805.4  &      \ion{Fe}{2}              & 2343.5 & 1.8 $\pm$ 0.2 & 1.477 \\ 
5108.5  & \ion{Zn}{2}/\ion{Cr}{2} blend & 2061.9 & 0.7 $\pm$ 0.3 & 1.478 \\ 
5018.5  & \ion{Zn}{2}/\ion{Mg}{1} blend & 2025.6 & 1.0 $\pm$ 0.3 & 1.478 \\ 
4478.9  &      \ion{Si}{2}              & 1808.0 & 0.5 $\pm$ 0.3 & 1.477 \\ 
4137.8  &      \ion{Al}{2}              & 1670.8 & 1.1 $\pm$ 0.3 & 1.477 \\ 
3838.0  &      \ion{C}{4} blend         & 1549.0 & 1.9 $\pm$ 0.4 & 1.478 \\ 
3781.8  &      \ion{Si}{2}              & 1526.7 & 1.4 $\pm$ 0.4 & 1.477 \\ 
\\
6045.0  &      \ion{Mg}{2}              & 2802.7 & 2.1 $\pm$ 0.2 & 1.157 \\ 
6028.2  &      \ion{Mg}{2}              & 2795.5 & 1.9 $\pm$ 0.2 & 1.156 \\ 
5606.7  &      \ion{Fe}{2}              & 2599.4 & 0.5 $\pm$ 0.3 & 1.157 \\ 
5574.1  &      \ion{Fe}{2}              & 2585.9 & 0.6 $\pm$ 0.3 & 1.156 \\ 
5137.2  &      \ion{Fe}{2}              & 2382.0 & 1.2 $\pm$ 0.3 & 1.157 \\ 
\\
5402.2  &      \ion{Mg}{2}?             & 2802.7 & 0.6 $\pm$ 0.3 & 0.927 \\ 
5389.1  &      \ion{Mg}{2}?             & 2795.5 & 0.9 $\pm$ 0.3 & 0.928 \\ 
\enddata
\end{deluxetable}

The redshift of \thisgrb\ is therefore unambiguously $z \geq 1.477$,
corresponding to the most distant absorber.  Furthermore, the
non-detection of Ly$\alpha$ forest absorption or continuum decrement
at $\lambda > 4000$ \AA\ would suggest the GRB host is at $z < 2.3$, though
this is not a firm constraint given the S/N ratio of the data. If the
highest-redshift lines are from the GRB host itself, then the GRB is
at $z = 1.477$. Several other GRB OTs have shown absorption line
systems which have been argued to arise from the ISM of the GRB host
galaxy: GRB~090508 \citep{metzger97}, GRB~980703 \citep{djorgovski98},
GRB~990123 \citep{kulkarni99}, GRB~990510 and GRB~990712
\citep{vreeswijk01}, and GRB~991216 \citep{vreeswijk99}.

While \ion{Mg}{2} absorption systems are commonly found along the
line-of-sight to distant QSOs, the three detected at $3800 < \lambda <
7500$~\AA---corresponding to a redshift interval of $0.36 < z < 1.68$,
or $\Delta z = 1.32$---is significantly larger than the mean of
$\langle N/z \rangle = 0.97$ \citep{steidel92}. Moreover, that value
was derived based on systems with a \ion{Mg}{2} $\lambda$2796 \AA\
rest-frame equivalent width $W_0(\lambda 2796) \geq 0.3$ \AA, measured
with generally similar spectral resolution as the observations
presented here. Though our detection threshhold for this line is
difficult to determine a priori, it is likely at least 0.6 \AA, and
\citet{steidel92} find an average of $\langle N/z \rangle = 0.52$ for
systems with such strong absorption ($W_0(\lambda 2796) \geq 0.6$
\AA). Thus, there is quite a discrepancy between the expectation of
$\sim$0.7 systems over the observed wavelength region compared to the
3 systems actually detected. This discrepancy would be mitigated if
the $z = 1.477$ system were the host galaxy of the GRB.

In addition, $z=1.477$ system has an unusually large rest-frame
equivalent width of the \ion{Mg}{2} $\lambda$ 2796 \AA\ line of
$W_0(\lambda 2796) = 3.0 \pm 0.2$~\AA, which is larger than all 111
systems found by \citet{steidel92}.  The $z=1.157$ system equivalent
width places it in the top 10\%, while the $z=0.928$ system is more
typical.  The \ion{Mg}{2} doublet ratios $W_0(\lambda
2796)/W_0(\lambda 2803)$ are $1.1\pm 0.3$, $0.9 \pm 0.3$, and $1.5 \pm
0.4)$ for systems with redshifts of 1.477, 1.157, and 0.928,
respectively; these values are typical of the anticorrelation between
$W(\lambda 2796)$ and the doublet ratio, indicating the lines are
strongly saturated \citep{steidel92}. The \ion{Fe}{2} and \ion{Mg}{1}
equivalent widths for the $z=1.477$ system are likewise unusually
strong \citep{churchill00}.  Since these absorption line strengths for
the $z=1.477$ are so large, either the impact parameter is extremely
small or this system represents the OT host galaxy.  Nearly all of the
absorption line systems in GRB OTs show substantially weaker lines
than the $z=1.477$ system \citep[the host of GRB~990712 is the lone
exception, with an equivalent width of $W_0(\lambda 2796)+W_0(\lambda
2803) = 9$~\AA;][]{vreeswijk01} further strengthening the case that
the high-redshift system is the OT host.

To make these arguments more quantitative, we calculate the Bayesian
odds ratio for the following two hypotheses (assumed to be equally
likely a priori): $H_1$, that the OT is at redshift $z > 1.477$ with
three foreground absorption systems, and $H_2$, that the OT is at
redshift $z = 1.477$ with two foreground absorption systems. For
$H_1$, the redshift path-length for detection of absorption systems is
our full spectral window,\footnote{Actually, this leads to a very
slight overestimate of the likelihood of $H_1$, because the spectral
window for detecting foreground \ion{Mg}{2} is less than our observed
wavelength range if the OT redshift is $z \leq 1.7$. This only
strengthens the argument presented.} $\Delta z = 1.32$, implying, as
above, an expectation of $\bar{N} = 0.52 \times 1.32 = 0.69$
absorption systems, assuming a detection threshold of $W_0 \geq 0.6$
\AA\ \citep{steidel92}. For $H_2$, the redshift path has an upper
bound at $z = 1.477$, yielding $\Delta z = 1.12$ and $\bar{N} = 0.52
\times 1.12 = 0.58$. The observed number of systems follows a simple
Poisson distribution, and so the Bayesian odds $p(H_2)/p(H_1)$ are
given by $[P(N=2;\bar{N}=0.58)/P(N=3;\bar{N}=0.69)] = 3.4$, indicating
that $H_2$ is more than three times as likely as $H_1$ under the given
assumptions. In addition to the \emph{number} of observed absorption
systems, we can also take into account their \emph{strength}, by
adopting the \ion{Mg}{2} $\lambda$2796 equivalent width distribution
function of \citet{steidel92}, $n(W_0) dW_0 \propto \exp(-W_0/W_0^{*})
dW_0$ with $W_0^{*}$ = 0.66 \AA\ (a good fit to the data for $W_0
\geq$ 0.6 \AA, our detection threshhold\footnote{\citet{steidel92}
note that for the strongest absorbers, $n(W_0)$ seems to change with
redshift, whereas we have assumed it to be fixed over the redshift
range $0.35 \leq z \leq 1.7$. However, we do not expect significant
changes in the results taking this evolution into account, because the
mean redshift of the absorber population that \citet{steidel92} used
to derive $n(W_0)$, $\langle z_{\rm abs}\rangle = 1.17$, well matches
our observable redshift range.}). Because of the rarity of absorption
systems as strong as the one detected at $z = 1.477$, incorporating
the line strengths into the calculation increases the Bayesian odds
favoring $H_2$ over $H_1$ to 28:1. Thus, there is strong support for
the identification of the $z = 1.477$ system as the host of
{\thisgrb}.

\begin{figure}
\epsscale{1.0}
\plotone{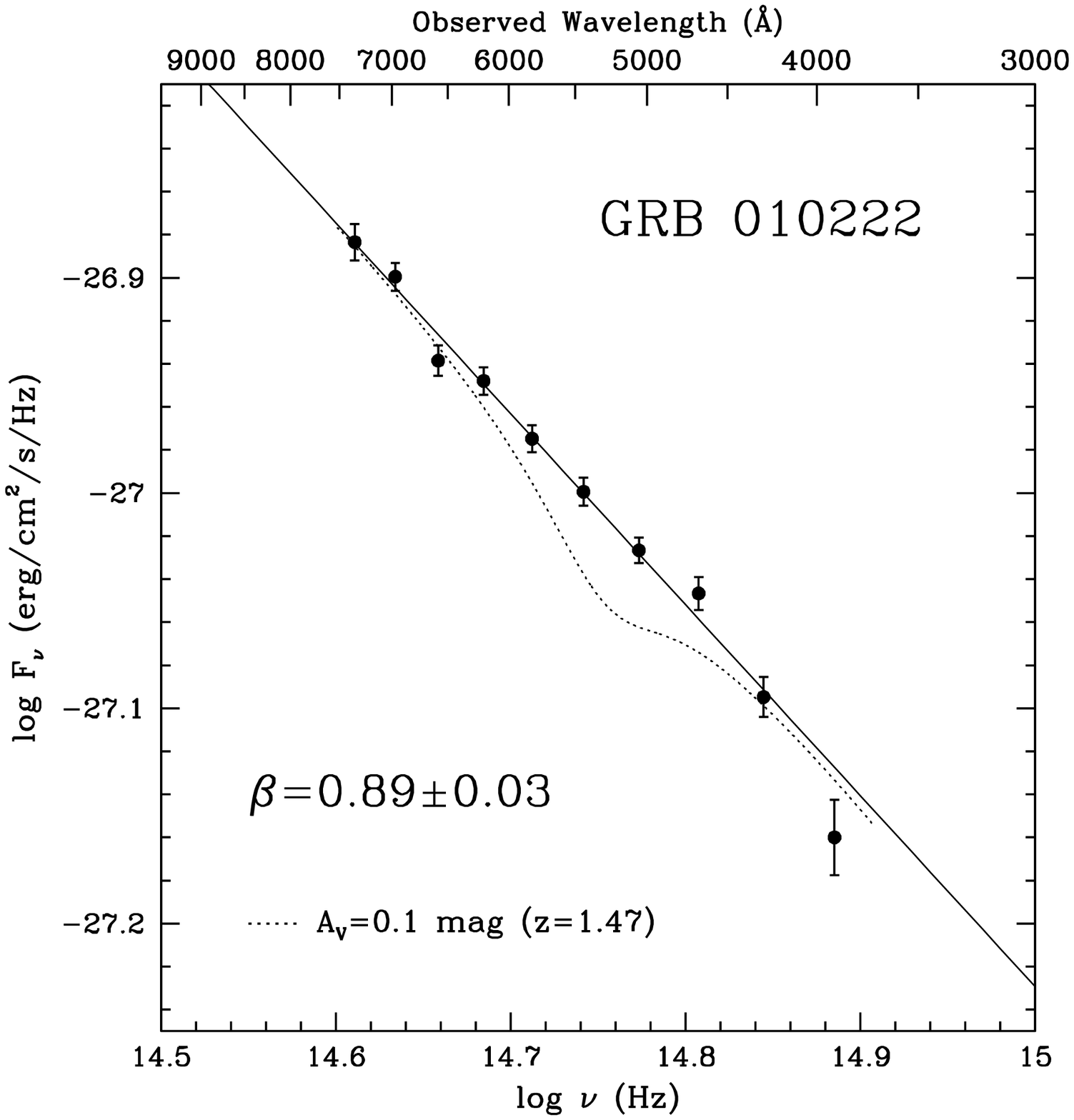}
\caption[Spectral continuum of \thisgrb]{Continuum flux of the OT
associated with \thisgrb\ based on the FLWO 1.5~m spectrum. The observed
flux has been corrected to match the FLWO 1.2~m photometry from the same
epoch \citep{stanek01}. The solid line shows the best-fit continuum
slope, $F_\nu \propto \nu^{-\beta}$, with $\beta = 0.89 \pm 0.03$. The
dotted line shows the effect of 0.1 mag of visual extinction from the
host galaxy, assuming an $R_V$ = 3.1 extinction law and $z = 1.477$ for
\thisgrb. The lack of a dust feature like the $\lambda$2175 \AA\ bump
suggests that either there is very little extinction from the $z =
1.477$ system or that the extinction law differs from the standard
Galactic extinction law. \label{fig-continuum}}
\end{figure}

Beyond the absorption systems along the line of sight, we also use the
observed spectrum to measure the spectral power-law index of the OT.
The continuum flux is overwhelmingly dominated by the OT, so
contamination from an underlying host or other objects on the slit is
negligible. After correcting for Galactic extinction, $E(\bv) = 0.023$
mag \citep*{schlegel98}, adopting a standard $R_V = 3.1$ extinction
law \citep*{cardelli89}, we bin the spectrum into ten segments
(excluding the absorption lines), using the rms deviation in each bin
as an estimate of the uncertainty, as shown in
Figure~\ref{fig-continuum}. We have adjusted the normalization to
match the concurrent photometry \citep{stanek01}. A least squares
minimization yields a power law index $\beta = 0.89 \pm 0.03$ (where
$F_\nu \propto \nu^{-\beta}$). As mentioned above, additional
uncertainty due to the relative flux calibration is likely to be
small. The spectral slope is rather steep in comparison to early
observations of other bright GRB afterglows.  For example GRB~990510
had a $\beta\simeq 0.5$ less than one day after the burst
\citep{stanek99} and GRB~991216 had $\beta\simeq 0.6$ after a
correction for large Galactic extinction \citep{garnavich00}. On the
other hand, GRB~000926 exhibited a very steep spectral slope with an
index of 1.5 which was attributed to significant dust extinction along
the line of sight \citep{price01}.

The steep spectral slope for \thisgrb\ may also indicate significant
extinction from dust in the host galaxy. Furthermore, jet models by
\citet*{sari99} predict a shallower spectral index than we observe
given the reported light curve decline rates \citep[for details,
see][]{stanek01}, also suggesting significant extinction. However, as
shown in Figure~\ref{fig-continuum}, the spectrum exhibits no clear
evidence for the $\lambda$2175~\AA\ ``bump'', typical of Galactic
interstellar dust \citep{cardelli89}, which falls in the observed
spectral region for $z=1.477$. Such a feature would be easily
detectable even at levels $A_V \simeq 0.1$ mag. Thus we conclude that
there is no significant extinction from Galactic-type dust in the $z =
1.477$ absorption system. However, substantial extinction from dust
with an extinction law like that found in the SMC could still account
for the steep spectral slope, as such dust does not show a significant
$\lambda$2175 \AA\ feature \citep{prevot84}. An SMC-like extinction
curve may be a more natural choice if GRBs come from young stellar
environments, since dust in starburst galaxies tends to lack the
$\lambda$2175 \AA\ bump \citep*{gcw97}.

\acknowledgments

We are grateful to the entire BeppoSAX team for the quick turnaround
in providing precise GRB positions to the astronomical community, as
well as to Scott Barthelmy and the GRB Coordinates Network (GCN). That
these results could be obtained with small aperture telescopes is
entirely due to the speed in which positions are reported and
disseminated. 







\begin{thebibliography}{}


\bibitem[Bloom et al.(2001a)Bloom, Kulkarni \& Djorgovski]{bloom01a} 
Bloom, J. S., Kulkarni, S. R., \& Djorgovski, S. G. 2001a, \aj, submitted
(astro-ph/0010176)

\bibitem[Bloom et al.(2001b)]{bloom01b} 
Bloom, J. S., Djorgovski, S. G., Halpern, J. P., Kulkarni, S. R.,
Galama, T. J., Price, P. A., \& Castro, S. M.  2001b, GCN Circular 989

\bibitem[Boella et al.(1997)]{bepposax} 
Boella, G., Butler, R. C., Perola, G. C., Piro, L., Scarsi, L., \&
Bleeker, J. A. M. 1997, \aaps, 122, 299

\bibitem[Bond(1997)]{bond97}
Bond, H. E. 1997, \iaucirc 6654

\bibitem[Cardelli et al.(1989)Cardelli, Clayton, \& Mathis]{cardelli89}
Cardelli, J. A., Clayton, G. C., \& Mathis, J. S. 1989, \apj, 345, 245

\bibitem[Castro et al.(2001)]{castro01}
Castro, S., et al. 2001, GCN Circular 999

\bibitem[Churchill et al.(2000)]{churchill00} 
Churchill, C. W., Mellon, R. R., Charlton, J. C., Jannuzi, B. T.,
Kirhakos, S., Steidel, C. C., \& Schneider, D. P. 2000, \apjs, 130,
91

\bibitem[Djorgovski et al.(1997)]{djorgovski97} 
Djorgovski, S. G., et al. 1997, \nat, 387, 876 

\bibitem[Djorgovski et al.(1998)]{djorgovski98} 
Djorgovski, S. G., Kulkarni, S. R., Bloom, J. S., Goodrich, R.,
Frail, D. A., Piro, L., \& Palazzi, E. 1998, \apjl, 508, L17

\bibitem[Fabricant et al.(1998)]{fast} 
Fabricant, D., Cheimets, P., Caldwell, N., \& Geary, J. 1998, \pasp, 110, 79 

\bibitem[Frail et al.(1997)]{frail97} 
Frail, D. A., Kulkarni, S. R., Nicastro, S. R., Feroci, M., \&
Taylor, G. B. 1997, \nat, 389, 261

\bibitem[Gandolfi(2001)]{gandolfi966}
Gandolfi, G. 2001, GCN Circular 966

\bibitem[Garnavich et al.(2000)]{garnavich00}
Garnavich, P. M., Jha, S., Pahre, M. A., Stanek, K. Z., Kirshner,
R. P., Garcia, M. R., Szentgyorgyi, A. H., \& Tonry, J. L. 2000,
\apj, 543, 61

\bibitem[Garnavich et al.(2001)]{garnavich965} 
Garnavich, P. M., Pahre, M. A., Jha, S., Calkins, M., Stanek, K. Z.,
McDowell, J., \& Kilgard, R. 2001, GCN Circular 965

\bibitem[Gordon et al.(1997)Gordon, Calzetti, \& Witt]{gcw97}
Gordon, K. D., Calzetti, D., \& Witt, A.N. 1997, \apj, 487, 625

\bibitem[Henden(2001a)]{henden961}
Henden, A. A. 2001a, GCN Circulars 961 and 962

\bibitem[Henden(2001b)]{henden987}
Henden, A. A. 2001b, GCN Circular 987

\bibitem[Horne(1986)]{horne86}
Horne, K. 1986, \pasp, 98, 609

\bibitem[Jha et al.(2001)]{jha974}
Jha, S., Matheson, T., Calkins, M., Pahre, M. A., Stanek, K. Z.,
McDowell, J., Kilgard, R., \& Garnavich, P. M. 2001, GCN Circular 974

\bibitem[Klebesadel et al.(1973)Klebesadel, Strong, \& Olson]{klebesadel73}
Klebesadel, R. W., Strong, I. B., \& Olson, R. A.  1973, \apj, 182, L85

\bibitem[Kulkarni et al.(1999)]{kulkarni99} 
Kulkarni, S. R., et al. 1999, \nat, 398, 389

\bibitem[Kulkarni et al.(2000)]{kulkarni00}
Kulkarni, S. R., et al. 2000, in Gamma-Ray Bursts: 5th Huntsville
Symposium, eds. R. M. Kippen, R. S. Mallozzi, \& G. J. Fishman (New
York: AIP), 277

\bibitem[Lamb(1995)]{lamb95}
Lamb, D. Q. 1995, \pasp, 107, 1152 

\bibitem[McDowell et al.(2001)]{mcdowell963}
McDowell, J., Kilgard, R., Garnavich, P. M., Stanek, K. Z., \& Jha, S.
2001, GCN Circular 963

\bibitem[Meegan et al.(1992)]{meegan92} 
Meegan, C. A., Fishman, G. J., Wilson, R. B., Horack, J. M.,
Brock, M. N., Paciesas, W. S., Pendleton, G. N., \& Kouveliotou,
C. 1992, \nat, 355, 143

\bibitem[Metzger et al.(1997)]{metzger97} 
Metzger, M. R., Djorgovski, S. G., Kulkarni, S. R., Steidel, C.\
C., Adelberger, K. L., Frail, D. A., Costa, E., \& Frontera, F.\
1997, \nat, 387, 878

\bibitem[Paczy\'nski(1995)]{paczynski95} 
Paczy\'nski, B. 1995, \pasp, 107, 1167

\bibitem[Piro(2001)]{piro959} 
Piro, L. 2001, GCN Circular 959

\bibitem[Pr\'evot et al.(1984)]{prevot84}
Pr\'evot, M. L., Lequeux, J., Maurice, E., Pr\'evot, L., \& Rocca-Volmerange,
B. 1984, \aap, 132, 389

\bibitem[Price et al.(2001)]{price01}
Price, P. A., et al. 2001, \apjl, 549, L7

\bibitem[Sari et al.(1999)Sari, Piran, \& Halpern]{sari99}
Sari, R., Piran, T., \& Halpern, J. P. 1999, \apjl, 519, L17

\bibitem[Schlegel et al.(1998)Schlegel, Finkbeiner, \& Davis]{schlegel98}
Schlegel, D. J., Finkbeiner, D. P., \& Davis, M. 1998, \apj, 500, 525

\bibitem[Stanek et al.(1999)]{stanek99}
Stanek, K. Z., Garnavich, P. M., Kaluzny, J., Pych, W., \& Thompson,
I. 1999, \apjl, 522, 39

\bibitem[Stanek et al.(2001)]{stanek01}
Stanek, K. Z., et al. 2001, in preparation

\bibitem[Steidel \& Sargent(1992)]{steidel92} 
Steidel, C. C. \& Sargent, W. L. W. 1992, \apjs, 80, 1 

\bibitem[Stone(1977)]{stone77}
Stone, R. P. S. 1977, \apj, 218, 767

\bibitem[Szentgyorgyi et al.(2001)]{szent01}
Szentgyorgyi, A. H., et al. 2001, in preparation

\bibitem[van Paradijs et al.(1997)]{paradijs97} 
van Paradijs, J., et al. 1997, \nat, 386, 686 

\bibitem[Vreeswijk et al.(1999)]{vreeswijk99}
Vreeswijk, P. M., Rol, E., Hjorth, J., Kouveliotou, C., Pian, E.,
Palazzi, E., Pedersen, H., Gorosabel, J., Castro-Tirado, A., \&
Greinder, J. 1999, GCN Circular 496

\bibitem[Vreeswijk et al.(2001)]{vreeswijk01} 
Vreeswijk, P. M., et al. 2001, \apj, 546, 672 

\bibitem[Wade \& Horne(1988)]{wade88}
Wade, R. A., \& Horne, K. D. 1988, \apj, 324, 411

\end{thebibliography}
\end{document}